\documentclass[proof]{he_xcop} 

\articletype{Original Paper}%

\received{2019}
\revised{2019}
\accepted{2019}

\usepackage{times,epsfig,amsmath} 
\usepackage{graphicx}

\newcommand{\planck}{{\it Planck}}
\newcommand{\xmm}{{\it XMM-Newton}}
\newcommand{\xmms}{{\it XMM}}
\newcommand{\kms}{km s$^{-1}$ Mpc$^{-1}$}

\begin{document}

\title{Helium abundance (and $H_0$) in X-COP galaxy clusters}

\author[1,2]{S. Ettori*}

\author[3]{V. Ghirardini}

\author[4]{D. Eckert}

\authormark{S. Ettori \textsc{et al}}

\address[1]{INAF, Osservatorio di Astrofisica e Scienza dello Spazio, via Pietro Gobetti 93/3, 40129 Bologna, Italy}
\address[2]{INFN, Sezione di Bologna, viale Berti Pichat 6/2, I-40127 Bologna, Italy}
\address[3]{Center for Astrophysics | Harvard \& Smithsonian, 60 Garden Street, Cambridge, MA 02138, USA}
\address[4]{Department of Astronomy, University of Geneva, Ch. d'Ecogia 16, CH-1290 Versoix, Switzerland}

\corres{*Corresponding author; \email{stefano.ettori@inaf.it}}


\abstract{We present the constraints on the helium abundance in 12 X-ray luminous galaxy clusters that have 
been mapped in their X-ray and Sunyaev-Zeldovich (SZ) signals out to $R_{200}$ for the \xmm\ Cluster Outskirts Project (X-COP). 
The unprecedented precision available for the estimate of $H_0$ allows
us to investigate how much the reconstructed X-ray and SZ signals are consistent with the expected ratio $x$
between helium and proton densities of 0.08--0.1.
We find that an $H_0$ around 70 \kms\ is preferred from our measurements, with lower 
values of $H_0$ as suggested from the Planck collaboration ($67$ \kms) requiring 
a 34\% higher value of $x$. 
On the other hand, higher values of $H_0$, as obtained by measurements in the local universe, 
impose an $x$, from the primordial nucleosynthesis calculations and current solar abundances,
reduced by 37--44\%. 
}




\maketitle


\section{Introduction}

Galaxy clusters form under the action of gravity by the accretion onto a dark matter halo of primordial gas, mostly composed by hydrogen and helium.
During its hierarchical assembly, this intracluster medium (ICM) is heated up to temperatures of $10^7-10^8$ K, making it 
an almost fully ionized plasma, which produces both X-ray emission through bremsstrahlung radiation and line emission, and a characteristic spectral
distortion of the cosmic microwave background (CMB) signal due to inverse Compton scattering off the hot ICM electrons of the CMB photons, the so-called
Sunyaev-Zel'dovich (SZ) effect \citep{SZ} detected at microwave wavelengths.

In the ICM, the helium is fully ionized and not directly observable, and can be different from the primordial amount due to, e.g.,
release from stars, or sedimentation, mostly in cluster cores under the action of the gravitational force.
if the suppression due to the magnetic field is modest, diffusion can occur and He (but not heavier metals like Fe) can drift inwards
with typical velocity \citep[e.g.][]{spitzer56,ettori06}
$v_{\rm sed} = A_1 A_2 v_{\rm th}^3 g / (Z_1^2  Z_2^2 n_1 \ln \Lambda) \approx$ 2 kpc / Gyr $\approx$ 2 km/s,
where labels `1' and `2' identify the two populations of particles with atomic weight $A$ and atomic number $Z$,
$v_{\rm th} = (2 kT /A_1 m_{\rm p})^{1/2}$ is the thermal velocity in a plasma with temperature $kT$ in hydrostatic equilibrium with 
gravitational acceleration $g$.
However, this effect can be significantly limited by the magnetic topology, plasma instabilities, gas mixing by mergers and turbulence.
For instance, $v_{\rm sed}$ is well below the expected level of turbulence present in the ICM of few hundred km/s
\citep[see e.g.][]{sanders13,hitomi_gasdyn18}, making even more inefficient the process to sediment, even partially, 
into the cluster core and within the Hubble time.
If helium sedimentation occurs, however, the expected impact is very limited:
an enhancement by 10\% of the helium abundance has been shown to affect only the metal
abundances by 0.02--0.03 over the Hitomi SXS band \citep[][]{hitomi18atomic}; 
on the integrated quantities, like gas and total mass, an effect is potentially measurable only in
very inner regions, and is negligible when large cluster's volume is considered 
\citep[$\sim R_{500}$\footnote{$R_{500}$ indicates the radius of the sphere enclosing an average mass density equal to 
500 times the critcal density of the Universe at the cluster's redshift}; see e.g.][]{bulbul11}.

In the X-ray spectral analysis, it is generally assumed that the helium abundance is equal to its primordial value.
At the present, calculations on the Big Bang Nucleosynthesis (BBN) provides rigid predictions on this primordial value, 
because complementary measurements on, e.g., the number of light neutrino families, the lifetime of the neutron, the baryonic density of the Universe 
are now constrained at percent level, leaving no more free parameter in standard BBN.
Hence, the calculated primordial abundances are in principle only affected by the moderate uncertainties in nuclear cross sections
\citep[see e.g.][]{pitrou18}. These calculations predict $Y_{\rm BBN} = \rho_{\rm He} / \rho_{\rm b} = 0.2471 \pm 0.0002$.

Observational constraints on pristine abundances refer to classes of objects believed to be more primitive, where the metallicity is expected 
to be less polluted from subsequent enrichment processes by massive stars.
For instance, $^4$He abundance is estimated in HII (ionized hydrogen) regions inside compact blue galaxies, assumed to be the constituents 
of present-day galaxies in our hierarchical structure formation paradigm.
By extrapolating the observed values in these metal poor regions to zero metallicity, $Y = \rho_{\rm He} / \rho_{\rm b}$ is measured to be
$0.2449\pm 0.0040$ \citep[e.g.][]{aver15}, well in agreement with $Y_{\rm BBN}$.

The helium abundance affects also the anisotropy in the cosmic microwave background at intermediate angular scales ($1000 \leq l \leq 3000$),
the so-called ``damping tail'' since the anisotropy power on these angular scales is damped by photon diffusion during recombination \citep{silk68}.
\cite{planck18-6} discusses limits on the estimates of $Y$, also accounting for the partial degeneration induced from the relativistic degrees of freedom.
These constraints are perfectly consistent with $Y_{\rm BBN}$, setting a scenario in which the primordial cosmic helium abundance is known 
at very high accuracy.

In this context, we investigate with the present study how the estimate of the He abundance affects 
the observed properties of the ICM, and which constraints we can put on it using 
the combination of the gas pressure profiles obtained for a well-selected sample of galaxy clusters with independent measurements of their
X-ray and SZ signals, and a knowledge on the value of the Hubble constant.

The paper is organized as follows. In the next Section, we describe the method adopted to constrain the He abundance, and the observables used.
We present our results in Section~3, drawing some conclusions in Section~4.
Unless mentioned otherwise, the quoted errors are statistical uncertainties at the $1 \sigma$ confidence level,
and the cosmological model of reference is a $\Lambda CDM$ with parameters
$H_0=$70 \kms\ and $\Omega_{\rm m} = 1 - \Omega_{\Lambda}=0.3$.

\section{He abundance in the ICM}

The helium affects 
the X-ray emission mainly through its contribution to the thermal bremsstrahlung \citep[see e.g.][]{qin00,ettori06,markevitch07}.
Following previous work, we define the ratio between the number densities of helium and protons
\begin{equation}
x = \frac{n_{\rm He}}{n_{\rm p}}.
\end{equation}

\begin{figure}[ht]
\centering
\includegraphics[page=2,trim=0 38 0 245,clip,width=\hsize]{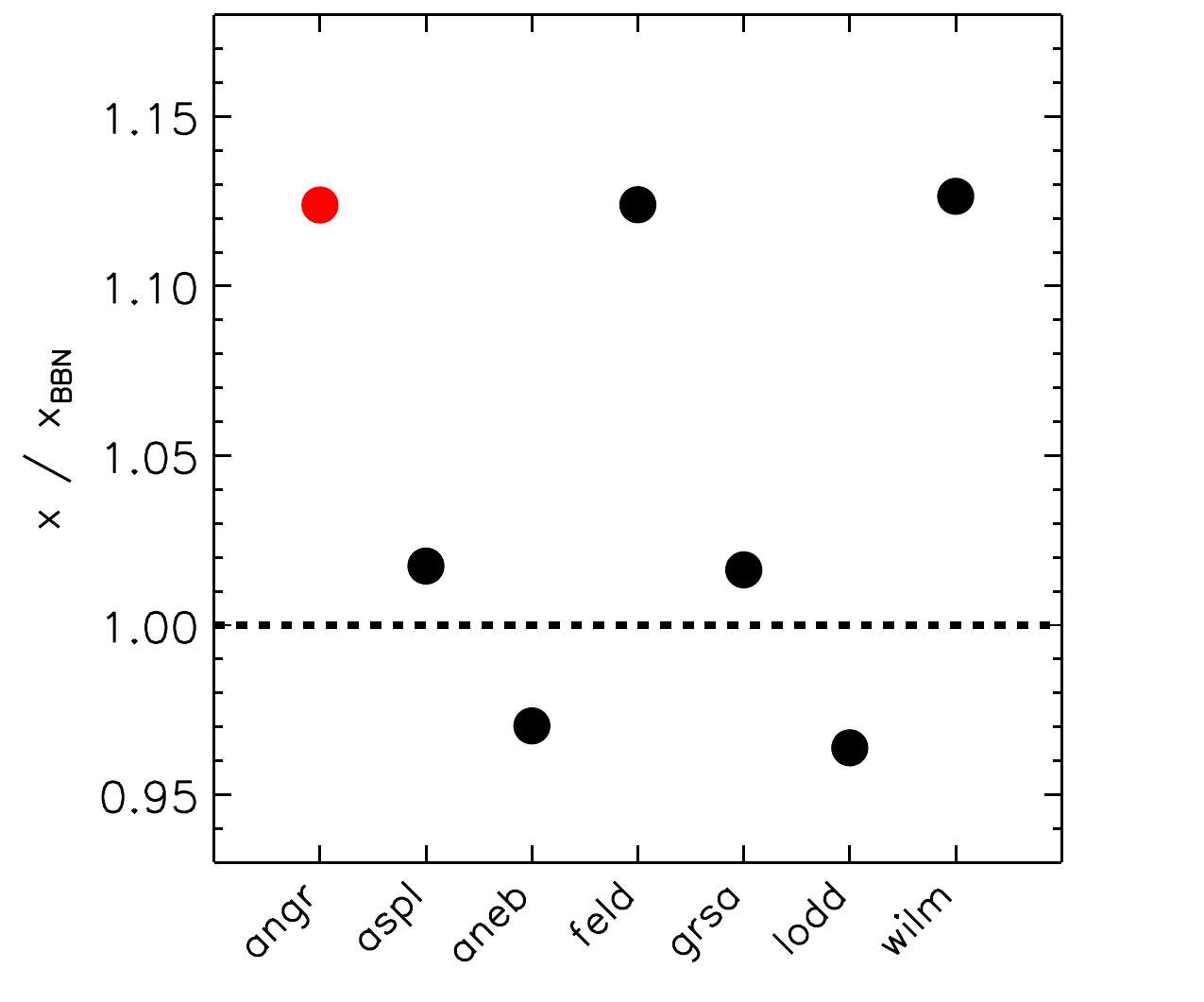} 
\caption{Comparison between the value of $x$ for each of the abundance tables available in 
{\tt Xspec} and listed in abscissa, and the primordial value $x_{\rm BBN}$.
}
\label{fig:he_bbn}
\end{figure}

In the X-ray analysis, 
$x$ is generally fixed to a value that will depend on the abundance table of reference.
In Fig.~\ref{fig:he_bbn}, we show the comparison between the values of $x$ tabulated in the abundance tables
available in the software adopted for our X-ray analysis \citep[{\tt Xspec},][]{xspec}\footnote{see https://heasarc.gsfc.nasa.gov/xanadu/xspec/manual/node117.html} and the one from primordial nucleosynthesis, $x_{\rm BBN} = (Y_{\rm BBN} / A_{\rm He})  \sum_i A_i n_i/n_{\rm H} = 0.0869$, where
we use $Y_{\rm BBN} = 0.2471$ (see previous Section) and the number densities, relative to hydrogen, tabulated in \cite{ag89} with a metallicity of 0.3 
(apart from H and He assumed to have a cosmic abundance).
We note the different behaviour of two popular tables: while {\it angr} \citep{ag89} has a value of $x$ (0.0977) about 12\% higher of $x_{BBN}$,
{\it aspl} \citep{aspl09} is well in agreement within 2\% (0.0851).

The X-ray brehmsstrahlung emissivity $\epsilon = n_{\rm e}  \sum(Z_i^2 n_i) \Lambda(T)$, where $n_{\rm e}$ is the electron number density and
$\Lambda(T)$ is the cooling function that depends only on the gas temperature $T$ \citep[less significantly moving to softer energy bands 
in the observer rest-frame; see e.g. fig.~2 in][]{ettori00}, can be well approximated as produced mainly from nuclei of hydrogen and helium ($Z_{\rm He} = 2$): 
$\epsilon \propto n_{\rm e} n_{\rm p} (1+4x) \sim n_{\rm p}^2  (1+4x)(1+2x)$, where the relation $n_{\rm e} / n_{\rm p} = 1 \,+2 x$ holds.
This approximation is reasonable because the contribution from other metals with atomic number $Z_i \ge 3$ raises the value of $(1+4x)(1+2x)$ by about 3\%.

Hence, for an observed X-ray flux $f \propto \epsilon \, R^3 / d_L^2$, $n_{\rm p}$ scales as $h^{0.5} \left[ (1+4x)(1+2x) \right]^{-0.5}$,
where $d_L$ is the luminosity distance that is proportional to the Hubble constant $h^{-1}$, and $R$ is the proper radius equal to the angular scale times the 
angular diameter distance $d_A = d_L / (1+z)^2$.

We can also write the dependence for the gas mass $M_{\rm gas}$ and the hydrostatic mass $M_{\rm hyd}$
\citep[see e.g.][]{ettori+13} as
\begin{eqnarray}
\mu & = & \left( \sum_i \frac{X_i n_i}{A_i} \right)^{-1} \sim \frac{1+4x}{2+3x},  \nonumber \\
M_{\rm gas} & \propto &  \mu (n_{\rm e} +n_{\rm p}) \, R^3 \sim \mu \, n_{\rm p} (2 +2x) \, R^3 \nonumber \\
 & \sim & \left( \frac{1+4x}{1+2x}\right)^{0.5} \frac{1+x}{2+3x} \, h^{-2.5},  \nonumber \\
M_{\rm hyd} & \propto &  \mu^{-1} \, R \sim \frac{2+3x}{1+4x} \, h^{-1},
\label{eq:corr}
\end{eqnarray}
where $\mu$ is the mean molecular weight, with $X_i$ and $A_i$ being the mass fraction and
atomic weight of element $i$, respectively. 

\begin{figure}[ht]
\centering
\includegraphics[page=9,trim=0 38 0 245,clip,width=\hsize]{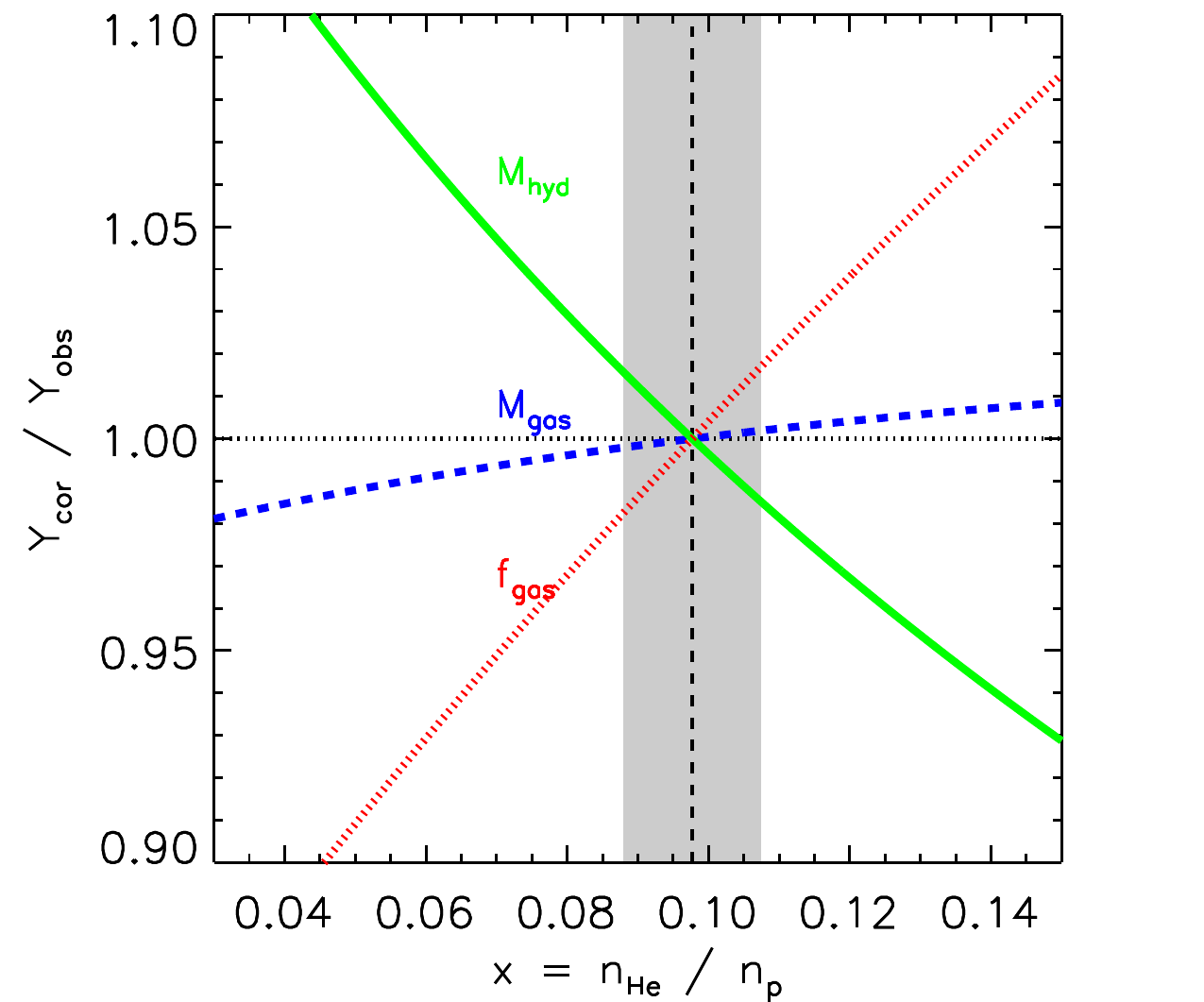} 
\caption{Variations with $x$ of $M_{\rm gas}$ (blue dashed line), $M_{\rm hyd}$ (green solid line) and $f_{\rm gas}  = M_{\rm gas}/M_{\rm hyd}$ (red dotted line) 
for a fixed value of $H_0$.
The vertical line indicates the helium abundance from {\it angr}, 
$x = 0.0977$, with $\pm$10\% represented by the shaded area.
}
\label{fig:cor_h0fix}
\end{figure}

In these equations, we make also evident the dependence upon the Hubble constant $h \equiv h_{70} = H_0 / ($70 \kms$)$, 
that is propagated through the radial dependence of these quantities and that will be useful to consider in the following analysis.
In general, for a fixed value of $H_0$, the impact of changing the helium abundance is in the order of few percent (see Fig.~\ref{fig:cor_h0fix}).

\subsection{Constraining $x$ with X-ray and SZ observations of the ICM}
\label{sect:method}

We discuss here how using the different dependence on $x= n_{\rm He} / n_{\rm p}$
of the ICM pressure recovered, independently, from X-ray and SZ signal can constrain the helium abundance.

The X-ray pressure is the product of the spectral measurement of the gas temperature, $T_{\rm e}$, by the electron density $n_{\rm e}$
estimated by the geometrical deprojection of the observed surface brightness $S_X \propto \int n_{\rm e} n_{\rm p} (1+4x) dl$, that implies the following scaling:
\begin{equation}
n_{\rm e, X} \sim {\rm deproj}(S_X)^{0.5}  \left( \frac{1+2x}{1+4x} \right)^{0.5} \, h^{0.5}.
\label{eq:nex}
\end{equation}
The SZ pressure is obtained directly from the deprojection of the Compton parameter $y \propto \int n_{\rm e} \, T_{\rm e}   \, dl$:
\begin{equation}
\left( n_{\rm e}  T_{\rm e} \right)_{SZ} \sim  {\rm deproj}(y) \, h.
\label{eq:nesz}
\end{equation}
In both cases, $dl$ indicates that the integration occurs along the light of sight and is thus proportional to the angular diameter distance and, hence, to $h^{-1}$.

\begin{figure}[t]
\centering
  \includegraphics[page=3,trim=0 38 0 245,clip,width=\hsize]{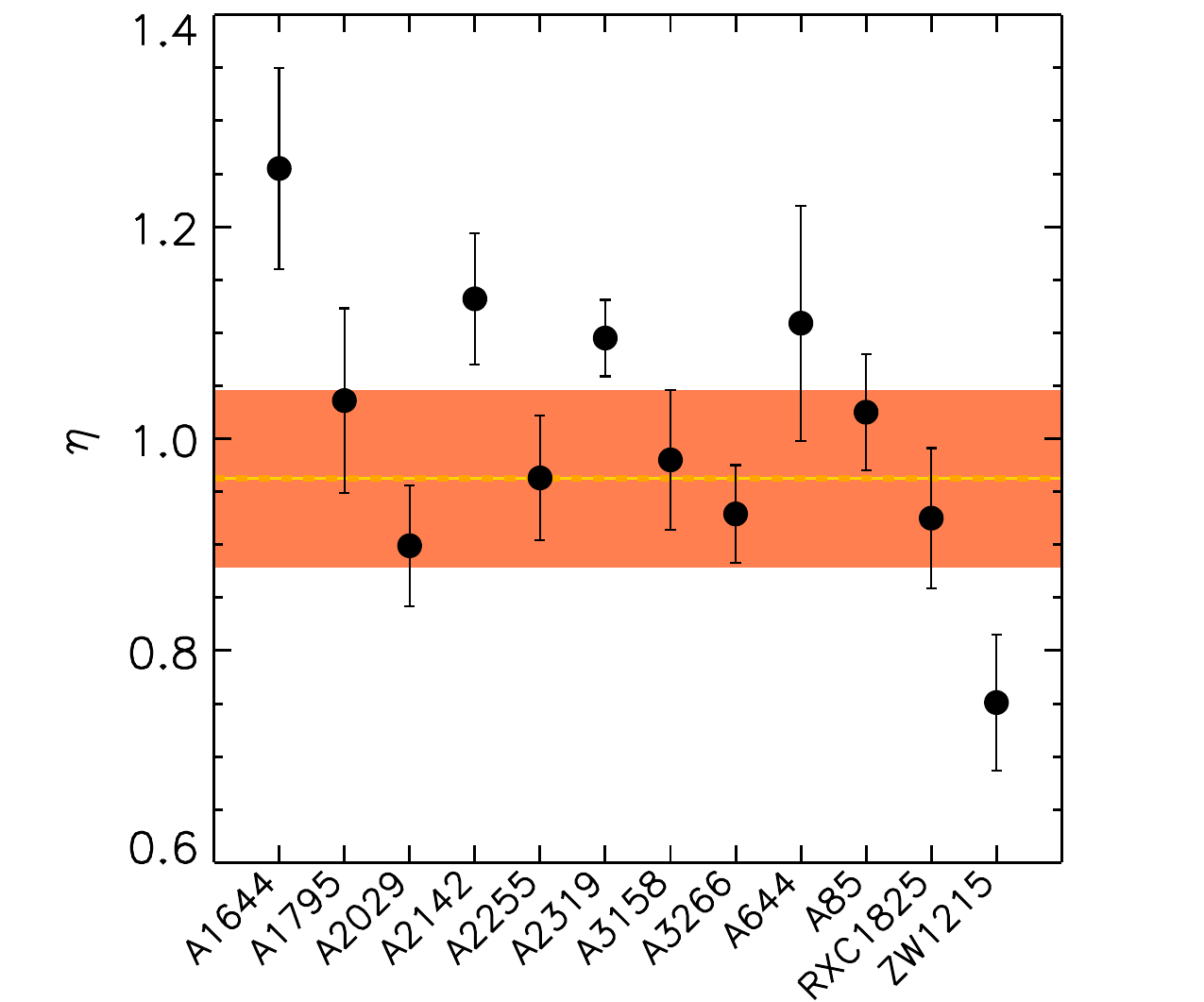} 
\caption{
Result on the parameter $\eta= P_{SZ}/P_{X}$ obtained from a best-fit of the pressure profiles for each X-COP object. 
The coloured areas indicate the scatter (0.0838) in the distribution and the error (0.0013) on the central position 
of the joint-fit value $\eta_{X-COP} = 0.9624$, used in the present analysis. See also appendix in \cite{ghi19univ}.
}
\label{fig:eta}
\end{figure}

Under the assumption of spherical symmetry, and that the gas density reconstructed from X-ray is not affected 
from clumpiness \citep[e.g.][]{nagai+11,roncarelli+13} that might bias high its value,
we can write the ratio between the two estimates of the pressure as
\begin{equation}
\frac{P_{SZ}}{P_X} = \frac{\left( n_{\rm e} \, T_{\rm e} \right)_{SZ}}{n_{\rm e, X} \, T_{\rm e}} \sim \left( \frac{1+4x}{1+2x} \right)^{0.5}  \, h^{0.5} = \eta.
\label{eq:eta}
\end{equation} 
The quantity $\eta = \theta^{0.5} \, h^{0.5}$, where  $\theta = (1+4x) / (1+2x)$, is then the one that we want to measure to constrain $x$ 
\citep[as originally suggested by][]{markevitch07},
also relying on independent measurements of the Hubble constant.
In the present work, we adopt the following values (with relative uncertainties) for the Hubble constant: 
$H_{0, CMB} = 67.4 \pm 0.5$ \kms\ from the \planck\ measurements of the CMB anisotropies, 
combining information from the temperature and polarization maps and the lensing reconstruction, assuming a spatially-flat 6-parameter $\Lambda$CDM cosmology \citep{planck18-6};
$H_{0, LMC} = 74.03 \pm 1.42$ \kms\  from Hubble Space Telescope observations of 70 long-period Cepheids in the Large Magellanic Cloud, combined 
with masers in NGC~4258, and Milky Way parallaxes \citep{riess19}.
These values embrace the actual extremes of the distribution of $H_0$. 
For example, other recent constraints are the one based on the tip of the Red Giant Branch \citep{freedman19}, 
that provides a value of $H_{0, TRGB} = 69.8$ \kms, and a measurement of $H_{0, delay} = 73.3$ \kms\ 
that relies on the joint analysis of six gravitationally lensed quasars with measured time delays \citep{wong19}, 
both with a relative uncertainty of $\sim$ 2.4--2.6 \% \citep[see further discussion in][]{verde19}.

It is worth noticing that, reversely, by assuming known $x$, the method can be used to constrain 
$H_0$ \citep[e.g.][]{cavaliere79,silk78} as discussed recently in an extensive way by \cite{kozmanyan19}.

\section{Constraints on $x$}
In this section, we present the sample of galaxy clusters that we have analyzed to recover both $P_X$ and $P_{SZ}$, and to constrain $\eta$ and hence $x=n_{\rm He} / n_{\rm p}$.

\subsection{The X-COP sample}
\label{subsect:xcop}

The   \xmm\ Cluster Outskirts Project \citep[X-COP\footnote{\url{https://www.astro.unige.ch/xcop}},][]{eck17xcop} is an \xmms\ Very Large Program dedicated to the study of the X-ray emission in cluster outskirts.
It has targeted 12 local, massive galaxy clusters selected for their high signal-to-noise ratios in the \planck\ all-sky SZ survey
\citep[S/N> 12 in the PSZ1 sample,][]{SZcatalog} as resolved sources ($R_{500} >$ 10 arcmin)
in the redshift range $0.04 < z < 0.1$ and along the directions with a galactic absorption lower than $10^{21}$ cm$^{-2}$ to avoid
any significant suppression of the X-ray emission in the soft band where most of the spatial analysis is performed.
These selection criteria guarantee that a joint analysis of the X-ray and SZ signals allows
the reconstruction of the ICM properties out to $R_{200} (\approx 1.5 R_{500})$ for all our targets. 
A complete description of the reduction and analysis of our proprietary X-ray data and of the \planck\ SZ data 
is provided in \cite{ghi19univ} \citep[see also][]{eckert19,ettori19}. 
Here, we want only to remark that a proper treatment of the X-ray surface brightness profiles, accounting for the median
of the distribution of the counts per pixel in a given radial annulus instead of the mean \citep[e.g.][]{zhu13,eckert+15}, guarantees
in the X-COP analysis against a relevant contribution from clumped gas that might systematically bias 
the estimate of the gas density (Eq.~\ref{eq:nex}). 

Finally, we have to initialize our measurement of $\eta$, by considering the conversion factors applied in our X-ray analysis, that 
is the only part in the calculation of $\eta$ where $x$ appears (see Eq.~\ref{eq:nex} and \ref{eq:nesz}).
As detailed in \cite{ghi19univ}, we consider a 30\% solar abundance metallicity, as in \cite{ag89}, to convert the emissivity 
$n_{\rm e} \sum(Z_i^2 n_i)$ to 1.7181 $n_{\rm p}^2$, and a fixed relation $n_{\rm e} = 1.17 n_{\rm p}$,
implying that $\eta = \eta_{X-COP} \, \left( 1.7181 / 1.17^2 \right)^{0.5}$,
where $\eta_{X-COP}  = 0.9624 \pm 0.0013$ (r.m.s. 0.0838; see Fig.~\ref{fig:eta}) 
is the joint best-fit of the SZ and X-ray pressure profiles recovered for the X-COP sample.
This joint fit mitigates any bias in the assumed spherical symmetry, in particular,  
of the SZ signal, where it is assumed that the line-of-sight gas distribution is the same as that in the plane of the sky.
In the following analysis, we consider the error on the central value as statistical uncertainty of $\eta$, 
whereas the dispersion around it indicates the limitations still present in our modelization of $\eta$ (see Sect.~\ref{sect:concl}).

\subsection{$\eta + H_0$: results on $x$}
\label{subsect:results}

\begin{figure}[ht]
\centering
\includegraphics[page=4,trim=0 38 0 245,clip,width=\hsize]{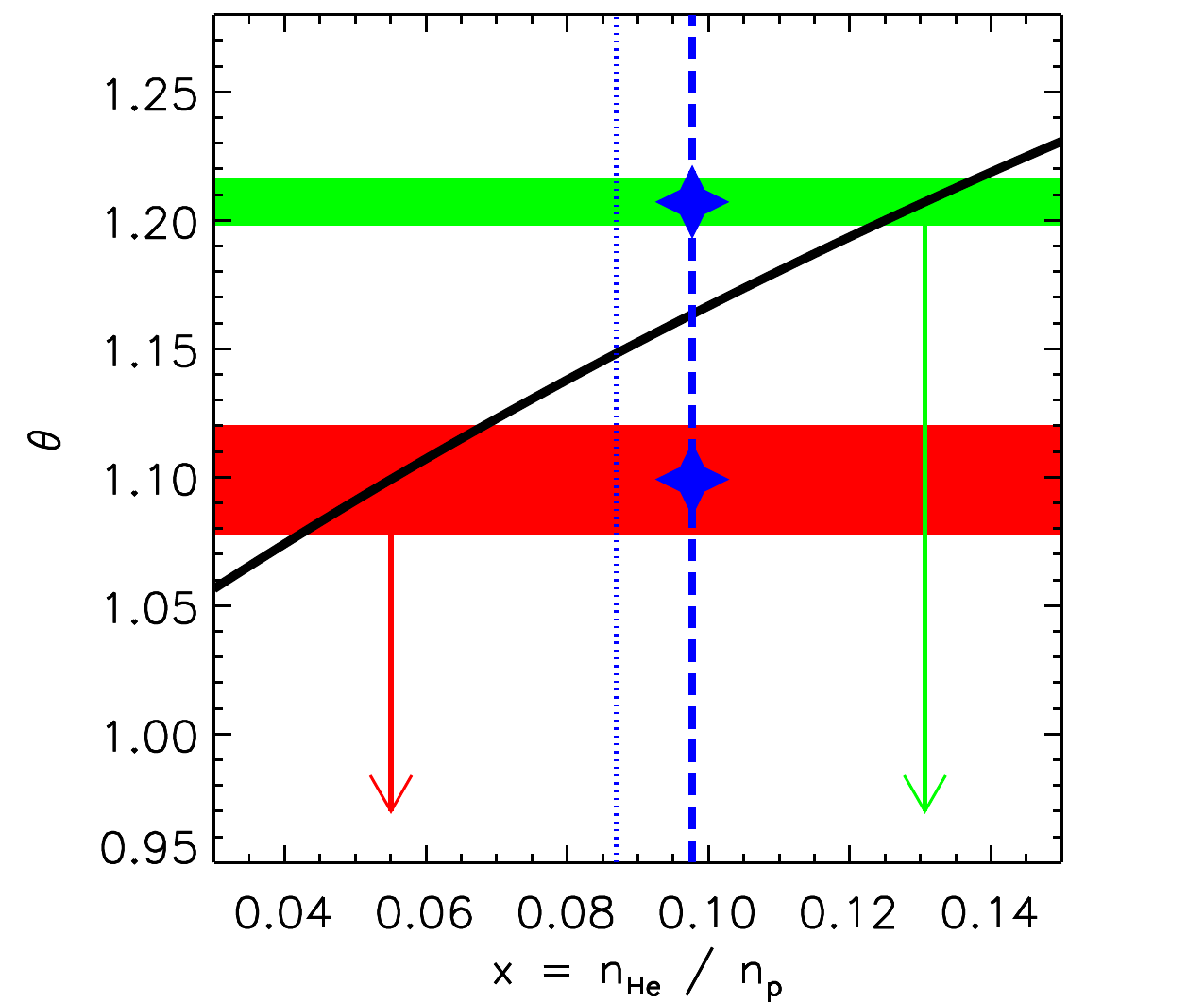} 
\caption{
Constraints on $\theta = \eta^2 / h$, and relative uncertainties at  $1 \sigma$, assuming $H_{0, CMB}$  (green area) and $H_{0, LMC}$ (red area).
The blue stars indicate the central value for the adopted value of $x$ in the X-COP X-ray analysis 
({\it angr}: dashed line; cosmological value: dotted line).
The black solid line represents $\theta =(1+4x) / (1+2x)$.
The coloured arrows indicate the preferred values of $x$ for a given $\theta$.
High values of $H_0$ tend to prefer low values of $x$.
}
\label{fig:theta}
\end{figure}

From the measurements of $\eta$ obtained in the X-COP sample and the adopted values of $H_0$ (see subsection~\ref{sect:method}), 
we can use Equation~\ref{eq:eta} to estimate $\theta = \eta^2 / h$.
We show the constraints on $\theta$ in Fig.~\ref{fig:theta}.
Low values of $x$, lower than the one ({\it angr}) adopted in the X-COP X-ray analysis seem to be preferred from high values of the Hubble constant.
In particular, for $H_{0, LMC}$ (74 \kms), $x = 0.055 \pm 0.013$, whereas for $H_{0, CMB}$ (67.4 \kms), $x = 0.131 \pm 0.008$. 
These values should be compared with a cosmic value of $x_{BBN} = 0.0869$ and $x =0.0977$ for the abundance table in {\it angr}. 

Reversely, fixing $x$ equal to the values from BBN, {\it aspl}, {\it angr}, we measure  
$H_0 = 70.9, 71.0, 69.9$ \kms, respectively, with a statistical error of 0.2 (12, when the scatter is considered) \kms .

\subsection{Corrections on the derived quantities}
\label{subsect:corr}

\begin{figure}[ht]
\centering
\includegraphics[page=8,trim=0 38 0 245,clip,width=\hsize]{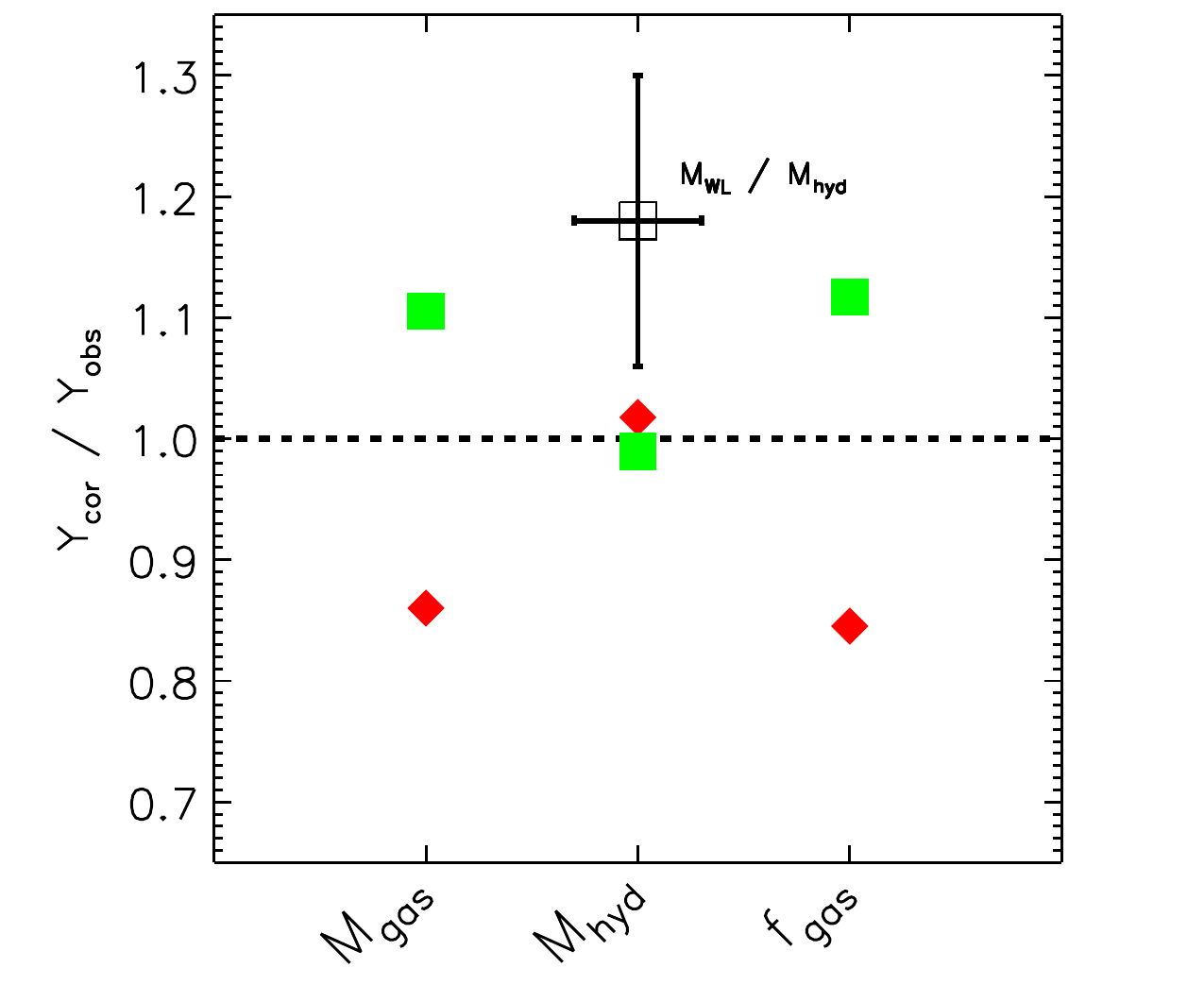} 
\caption{Total corrections  on  $M_{\rm gas}$, $M_{\rm hyd}$ and $f_{\rm gas}$, assuming $H_{0, CMB}$  (green squares) and $H_{0, LMC}$ (red diamonds).
The correction on the hydrostatic mass is compared with the measured tension with weak-lensing estimates for a subsample of 6 X-COP objects 
\citep[for details, see][]{ettori19}.
}
\label{fig:cor}
\end{figure}

From Fig.~\ref{fig:theta}, we can derive, for each assumed value of $H_0$, the expected $\bar{\theta}$ and, then, the corresponding $\bar{x}$.
Thus, we can associate to a given $H_0$ the correction which propagates to the gas mass and to the hydrostatic mass through $x$, accordingly to the
scaling presented in Eq.~\ref{eq:corr} and accounting for the dependence both on $x$ and on $H_0$. 
We show in Fig.~\ref{fig:cor} the total correction on these quantities.
Overall, the corrections should be less than $\sim15\%$, with lower tension (below 10\%) obtained with $H_{0, CMB}$.
Note that the estimated difference with respect to the reconstructed value of $M_{\rm hyd}$ is, in both cases, of few per cent and 
well below the hydrostatic bias measured by using weak lensing mass estimates.

\begin{figure}[ht]
\centering
\includegraphics[page=10,trim=0 38 0 245,clip,width=\hsize]{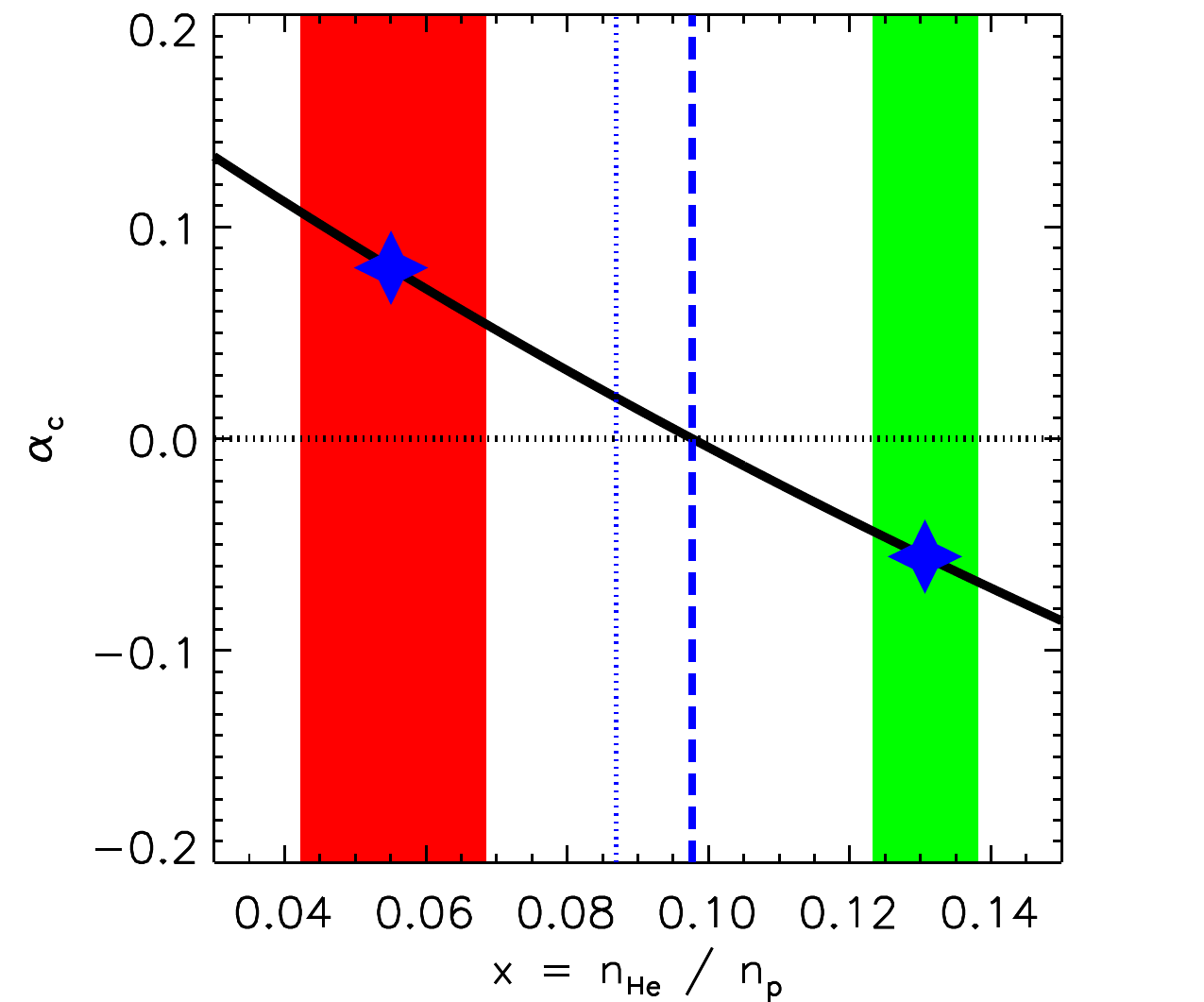} 
\caption{Constraints on $\alpha_c$ (see Eq.~\ref{eq:ac}). As in Fig.~\ref{fig:theta}, the assumed $H_0$ is colour-coded (green: $H_{0, CMB}$; red: $H_{0, LMC}$).
The blue stars indicate the central value for $x$, to be compared with the value adopted in the X-ray analysis ({\it angr}, dashed line) and
the cosmological value (dotted line).
} \label{fig:ac}
\end{figure}

We present also the correction propagated to the gas mass fraction $f_{\rm gas} = M_{\rm gas}/M_{\rm hyd}$ through the scaling
presented in Eq.~\ref{eq:corr}.
We can write this correction in a way similar to the form adopted to represent how the fraction $\alpha_P$ of the non-thermal pressure 
with respect to the total one (and equivalent to the hydrostatic bias $b$, when $\alpha_P$ does not vary with the radius) 
propagates into the estimate of $f_{\rm gas}$
\citep[see e.g. Eq.~8 in][]{eckert19}:
\begin{equation}
\frac{f_{\rm gas, true}}{f_{\rm gas, obs}} = \frac{ f_{\rm gas, true} }{f_{\rm gas, cor}} \frac{ f_{\rm gas, cor} }{f_{\rm gas, obs}}
       = (1 - \alpha_P) \, (1 - \alpha_c),
\label{eq:ac}
\end{equation}
where $f_{\rm gas, obs}$ is the observed gas fraction, $f_{\rm gas, true}$ is the expected ``true'' gas fraction, 
and $f_{\rm gas, cor}$ is the gas fraction after the correction by its dependence on the quantity $x$ and the Hubble constant $h$
(from $M_{\rm gas}$ and $M_{\rm hyd}$ in Eq.~\ref{eq:corr}).
In Fig.~\ref{fig:ac}, we show the constraints we obtain on $\alpha_c$.
Once again, the expected (either cosmological or from the adopted solar adundance) helium density lies between the amount required from 
the considered $H_0$, with values of $x$ below the canonical range of 0.08--0.1 corresponding to  $H_0$ above 70 \kms.

\section{Conclusions}
\label{sect:concl}

We have discussed the role of the helium abundance on the observed properties of the ICM in a well-selected sample of nearby massive 
galaxy clusters with independent X-ray and SZ measurements.

We obtain that the present constraints on the Hubble constant require central values of $x = n_{\rm He} / n_{\rm p}$ between 0.055
(for $H_{0, LMC}$ = 74 \kms) and 0.131 (for $H_{0, CMB}$ = 67.4 \kms), that encompass the assumed value 
in the X-ray analysis \citep[$x =0.0977$ for the abundance table in][]{ag89} and the cosmological one inferred from BBN ($x_{\rm BBN} = 0.0869$).
In any case, values of $H_0$ around 70 \kms\ are preferred from our estimates of $\eta = P_{SZ}/P_X = \theta^{0.5} \, h^{0.5}$, with $\theta = (1+4x) / (1+2x)$,
if $x$ has to lie around values of 0.08--0.1. 
While higher values of $x$, requiring lower estimates of $H_0$, can be partially explained with an efficient process of sedimentation, 
a reduction of the helium abundance (implying $H_0 > 70$ \kms) can be obtained by the action of thermal diffusion \cite[e.g.][]{medvedev14}. 
On the other hand, magnetic fields and mixing effects of large-scale turbulence are known to play a role in shaping the ICM properties, 
also by reducing significantly the impact of the above mentioned processes.
On the contrary, fixing $x$ to the values from BBN, \cite{aspl09} and \cite{ag89}, we obtain $H_0 = 70.9, 71.0, 69.9$ 
(1 $\sigma$ error: 0.2; 12, when the scatter in the distribution of $\eta$ -see Fig.~\ref{fig:eta}- is propagated) \kms, respectively.

A further improvement both on the statistical and systematic uncertainty of the estimate of $\eta$ will be obtained
with larger samples of X-ray and SZ estimates of the ICM pressure with respect to the dataset of 12 values from the X-COP sample.
One of these samples will be obtained by our dedicated \xmm\ Heritage program \footnote{\url{http://xmm-heritage.oas.inaf.it/}}, 
that will enlarge by a factor of $\sim$10 the number of measurements of $\eta$, allowing to reduce by $\sqrt{10} \sim 3.2$ 
the statistical error and to lower more significantly the impact of the assumed spherical geometry of the ICM distribution.


\section*{Acknowledgments}
The research leading to these results has received funding from the European Union's Horizon 2020 Programme under the AHEAD project (grant agreement n. 654215).
S.E. acknowledges financial contribution from the contracts ASI 2015-046-R.0 and ASI-INAF n.2017-14-H.0.






\small
\bibliography{he_xcop}

\begin{thebibliography}{}

\bibitem [\protect \citeauthoryear {%
{Anders}%
\ \BBA {} {Grevesse}%
}{%
{Anders}%
\ \BBA {} {Grevesse}%
}{%
{\protect \APACyear {1989}}%
}]{%
ag89}
\APACinsertmetastar {%
ag89}%
\begin{APACrefauthors}%
{Anders}, E.%
\BCBT {}\ \BBA {} {Grevesse}, N.%
\end{APACrefauthors}%
\unskip\
\newblock
\APACrefYearMonthDay{1989}{{\APACmonth{01}}}{},
\newblock
\unskip
\newblock
\APACjournalVolNumPages{\gca}{53}{}{197-214}.
\newblock
\begin{APACrefDOI} \doi{10.1016/0016-7037(89)90286-X} \end{APACrefDOI}
\PrintBackRefs{\CurrentBib}

\bibitem [\protect \citeauthoryear {%
{Arnaud}%
}{%
{Arnaud}%
}{%
{\protect \APACyear {1996}}%
}]{%
xspec}
\APACinsertmetastar {%
xspec}%
\begin{APACrefauthors}%
{Arnaud}, K\BPBI A.%
\end{APACrefauthors}%
\unskip\
\newblock
\APACrefYearMonthDay{1996}{}{},
\newblock
{\BBOQ}\APACrefatitle {{XSPEC: The First Ten Years}} {{XSPEC: The First Ten
  Years}}.{\BBCQ}
\newblock
\BIn{} G\BPBI H.~{Jacoby}\ \BBA {} J.~{Barnes}\ (\BEDS), \APACrefbtitle
  {Astronomical Data Analysis Software and Systems V} {Astronomical Data
  Analysis Software and Systems V}\ \BVOL~101, \BPG~17.
\PrintBackRefs{\CurrentBib}

\bibitem [\protect \citeauthoryear {%
{Asplund}%
, {Grevesse}%
, {Sauval}%
\BCBL {}\ \BBA {} {Scott}%
}{%
{Asplund}%
\ \protect \BOthers {.}}{%
{\protect \APACyear {2009}}%
}]{%
aspl09}
\APACinsertmetastar {%
aspl09}%
\begin{APACrefauthors}%
{Asplund}, M.%
, {Grevesse}, N.%
, {Sauval}, A\BPBI J.%
\BCBL {}\ \BBA {} {Scott}, P.%
\end{APACrefauthors}%
\unskip\
\newblock
\APACrefYearMonthDay{2009}{Sep}{},
\newblock
\unskip
\newblock
\APACjournalVolNumPages{\araa}{47}{1}{481-522}.
\newblock
\begin{APACrefDOI} \doi{10.1146/annurev.astro.46.060407.145222}
  \end{APACrefDOI}
\PrintBackRefs{\CurrentBib}

\bibitem [\protect \citeauthoryear {%
{Aver}%
, {Olive}%
\BCBL {}\ \BBA {} {Skillman}%
}{%
{Aver}%
\ \protect \BOthers {.}}{%
{\protect \APACyear {2015}}%
}]{%
aver15}
\APACinsertmetastar {%
aver15}%
\begin{APACrefauthors}%
{Aver}, E.%
, {Olive}, K\BPBI A.%
\BCBL {}\ \BBA {} {Skillman}, E\BPBI D.%
\end{APACrefauthors}%
\unskip\
\newblock
\APACrefYearMonthDay{2015}{Jul}{},
\newblock
\unskip
\newblock
\APACjournalVolNumPages{\jcap}{2015}{7}{011}.
\newblock
\begin{APACrefDOI} \doi{10.1088/1475-7516/2015/07/011} \end{APACrefDOI}
\PrintBackRefs{\CurrentBib}

\bibitem [\protect \citeauthoryear {%
{Bulbul}%
\ \protect \BOthers {.}}{%
{Bulbul}%
\ \protect \BOthers {.}}{%
{\protect \APACyear {2011}}%
}]{%
bulbul11}
\APACinsertmetastar {%
bulbul11}%
\begin{APACrefauthors}%
{Bulbul}, G\BPBI E.%
, {Hasler}, N.%
, {Bonamente}, M.%
, {Joy}, M.%
, {Marrone}, D.%
, {Miller}, A.%
\BCBL {}\ \BBA {} {Mroczkowski}, T.%
\end{APACrefauthors}%
\unskip\
\newblock
\APACrefYearMonthDay{2011}{Sep}{},
\newblock
\unskip
\newblock
\APACjournalVolNumPages{\aap}{533}{}{A6}.
\newblock
\begin{APACrefDOI} \doi{10.1051/0004-6361/201016407} \end{APACrefDOI}
\PrintBackRefs{\CurrentBib}

\bibitem [\protect \citeauthoryear {%
{Cavaliere}%
, {Danese}%
\BCBL {}\ \BBA {} {de Zotti}%
}{%
{Cavaliere}%
\ \protect \BOthers {.}}{%
{\protect \APACyear {1979}}%
}]{%
cavaliere79}
\APACinsertmetastar {%
cavaliere79}%
\begin{APACrefauthors}%
{Cavaliere}, A.%
, {Danese}, L.%
\BCBL {}\ \BBA {} {de Zotti}, G.%
\end{APACrefauthors}%
\unskip\
\newblock
\APACrefYearMonthDay{1979}{Jun}{},
\newblock
\unskip
\newblock
\APACjournalVolNumPages{\aap}{75}{3}{322-325}.
\PrintBackRefs{\CurrentBib}

\bibitem [\protect \citeauthoryear {%
{Eckert}%
\ \protect \BOthers {.}}{%
{Eckert}%
\ \protect \BOthers {.}}{%
{\protect \APACyear {2017}}%
}]{%
eck17xcop}
\APACinsertmetastar {%
eck17xcop}%
\begin{APACrefauthors}%
{Eckert}, D.%
, {Ettori}, S.%
, {Pointecouteau}, E.%
, {Molendi}, S.%
, {Paltani}, S.%
\BCBL {}\ \BBA {} {Tchernin}, C.%
\end{APACrefauthors}%
\unskip\
\newblock
\APACrefYearMonthDay{2017}{{\APACmonth{03}}}{},
\newblock
\unskip
\newblock
\APACjournalVolNumPages{Astronomische Nachrichten}{338}{}{293-298}.
\newblock
\begin{APACrefDOI} \doi{10.1002/asna.201713345} \end{APACrefDOI}
\PrintBackRefs{\CurrentBib}

\bibitem [\protect \citeauthoryear {%
{Eckert}%
\ \protect \BOthers {.}}{%
{Eckert}%
\ \protect \BOthers {.}}{%
{\protect \APACyear {2019}}%
}]{%
eckert19}
\APACinsertmetastar {%
eckert19}%
\begin{APACrefauthors}%
{Eckert}, D.%
, {Ghirardini}, V.%
, {Ettori}, S.%
\ et al.\end{APACrefauthors}%
\unskip\
\newblock
\APACrefYearMonthDay{2019}{Jan}{},
\newblock
\unskip
\newblock
\APACjournalVolNumPages{\aap}{621}{}{A40}.
\newblock
\begin{APACrefDOI} \doi{10.1051/0004-6361/201833324} \end{APACrefDOI}
\PrintBackRefs{\CurrentBib}

\bibitem [\protect \citeauthoryear {%
{Eckert}%
\ \protect \BOthers {.}}{%
{Eckert}%
\ \protect \BOthers {.}}{%
{\protect \APACyear {2015}}%
}]{%
eckert+15}
\APACinsertmetastar {%
eckert+15}%
\begin{APACrefauthors}%
{Eckert}, D.%
, {Roncarelli}, M.%
, {Ettori}, S.%
, {Molendi}, S.%
, {Vazza}, F.%
, {Gastaldello}, F.%
\BCBL {}\ \BBA {} {Rossetti}, M.%
\end{APACrefauthors}%
\unskip\
\newblock
\APACrefYearMonthDay{2015}{{\APACmonth{03}}}{},
\newblock
\unskip
\newblock
\APACjournalVolNumPages{\mnras}{447}{}{2198-2208}.
\newblock
\begin{APACrefDOI} \doi{10.1093/mnras/stu2590} \end{APACrefDOI}
\PrintBackRefs{\CurrentBib}

\bibitem [\protect \citeauthoryear {%
{Ettori}%
}{%
{Ettori}%
}{%
{\protect \APACyear {2000}}%
}]{%
ettori00}
\APACinsertmetastar {%
ettori00}%
\begin{APACrefauthors}%
{Ettori}, S.%
\end{APACrefauthors}%
\unskip\
\newblock
\APACrefYearMonthDay{2000}{Jan}{},
\newblock
\unskip
\newblock
\APACjournalVolNumPages{\mnras}{311}{2}{313-316}.
\newblock
\begin{APACrefDOI} \doi{10.1046/j.1365-8711.2000.03037.x} \end{APACrefDOI}
\PrintBackRefs{\CurrentBib}

\bibitem [\protect \citeauthoryear {%
{Ettori}%
\ \protect \BOthers {.}}{%
{Ettori}%
\ \protect \BOthers {.}}{%
{\protect \APACyear {2013}}%
}]{%
ettori+13}
\APACinsertmetastar {%
ettori+13}%
\begin{APACrefauthors}%
{Ettori}, S.%
, {Donnarumma}, A.%
, {Pointecouteau}, E.%
, {Reiprich}, T\BPBI H.%
, {Giodini}, S.%
, {Lovisari}, L.%
\BCBL {}\ \BBA {} {Schmidt}, R\BPBI W.%
\end{APACrefauthors}%
\unskip\
\newblock
\APACrefYearMonthDay{2013}{{\APACmonth{08}}}{},
\newblock
\unskip
\newblock
\APACjournalVolNumPages{\ssr}{177}{}{119-154}.
\newblock
\begin{APACrefDOI} \doi{10.1007/s11214-013-9976-7} \end{APACrefDOI}
\PrintBackRefs{\CurrentBib}

\bibitem [\protect \citeauthoryear {%
{Ettori}%
\ \BBA {} {Fabian}%
}{%
{Ettori}%
\ \BBA {} {Fabian}%
}{%
{\protect \APACyear {2006}}%
}]{%
ettori06}
\APACinsertmetastar {%
ettori06}%
\begin{APACrefauthors}%
{Ettori}, S.%
\BCBT {}\ \BBA {} {Fabian}, A\BPBI C.%
\end{APACrefauthors}%
\unskip\
\newblock
\APACrefYearMonthDay{2006}{Jun}{},
\newblock
\unskip
\newblock
\APACjournalVolNumPages{\mnras}{369}{1}{L42-L46}.
\newblock
\begin{APACrefDOI} \doi{10.1111/j.1745-3933.2006.00170.x} \end{APACrefDOI}
\PrintBackRefs{\CurrentBib}

\bibitem [\protect \citeauthoryear {%
{Ettori}%
\ \protect \BOthers {.}}{%
{Ettori}%
\ \protect \BOthers {.}}{%
{\protect \APACyear {2019}}%
}]{%
ettori19}
\APACinsertmetastar {%
ettori19}%
\begin{APACrefauthors}%
{Ettori}, S.%
, {Ghirardini}, V.%
, {Eckert}, D.%
\ et al.\end{APACrefauthors}%
\unskip\
\newblock
\APACrefYearMonthDay{2019}{Jan}{},
\newblock
\unskip
\newblock
\APACjournalVolNumPages{\aap}{621}{}{A39}.
\newblock
\begin{APACrefDOI} \doi{10.1051/0004-6361/201833323} \end{APACrefDOI}
\PrintBackRefs{\CurrentBib}

\bibitem [\protect \citeauthoryear {%
{Freedman}%
\ \protect \BOthers {.}}{%
{Freedman}%
\ \protect \BOthers {.}}{%
{\protect \APACyear {2019}}%
}]{%
freedman19}
\APACinsertmetastar {%
freedman19}%
\begin{APACrefauthors}%
{Freedman}, W\BPBI L.%
, {Madore}, B\BPBI F.%
, {Hatt}, D.%
\ et al.\end{APACrefauthors}%
\unskip\
\newblock
\APACrefYearMonthDay{2019}{Jul}{},
\newblock
\unskip
\newblock
\APACjournalVolNumPages{arXiv e-prints}{}{}{arXiv:1907.05922}.
\PrintBackRefs{\CurrentBib}

\bibitem [\protect \citeauthoryear {%
{Ghirardini}%
\ \protect \BOthers {.}}{%
{Ghirardini}%
\ \protect \BOthers {.}}{%
{\protect \APACyear {2019}}%
}]{%
ghi19univ}
\APACinsertmetastar {%
ghi19univ}%
\begin{APACrefauthors}%
{Ghirardini}, V.%
, {Eckert}, D.%
, {Ettori}, S.%
\ et al.\end{APACrefauthors}%
\unskip\
\newblock
\APACrefYearMonthDay{2019}{Jan}{},
\newblock
\unskip
\newblock
\APACjournalVolNumPages{\aap}{621}{}{A41}.
\newblock
\begin{APACrefDOI} \doi{10.1051/0004-6361/201833325} \end{APACrefDOI}
\PrintBackRefs{\CurrentBib}

\bibitem [\protect \citeauthoryear {%
{Hitomi Collaboration}%
\ \protect \BOthers {.}}{%
{Hitomi Collaboration}%
\ \protect \BOthers {.}}{%
{\protect \APACyear {2018}}%
{\protect \APACexlab {{\protect \BCnt {1}}}}}]{%
hitomi_gasdyn18}
\APACinsertmetastar {%
hitomi_gasdyn18}%
\begin{APACrefauthors}%
{Hitomi Collaboration}%
, {Aharonian}, F.%
, {Akamatsu}, H.%
\ et al.\end{APACrefauthors}%
\unskip\
\newblock
\APACrefYearMonthDay{2018{\protect \BCnt {1}}}{Mar}{},
\newblock
\unskip
\newblock
\APACjournalVolNumPages{\pasj}{70}{2}{9}.
\newblock
\begin{APACrefDOI} \doi{10.1093/pasj/psx138} \end{APACrefDOI}
\PrintBackRefs{\CurrentBib}

\bibitem [\protect \citeauthoryear {%
{Hitomi Collaboration}%
\ \protect \BOthers {.}}{%
{Hitomi Collaboration}%
\ \protect \BOthers {.}}{%
{\protect \APACyear {2018}}%
{\protect \APACexlab {{\protect \BCnt {2}}}}}]{%
hitomi18atomic}
\APACinsertmetastar {%
hitomi18atomic}%
\begin{APACrefauthors}%
{Hitomi Collaboration}%
, {Aharonian}, F.%
, {Akamatsu}, H.%
\ et al.\end{APACrefauthors}%
\unskip\
\newblock
\APACrefYearMonthDay{2018{\protect \BCnt {2}}}{Mar}{},
\newblock
\unskip
\newblock
\APACjournalVolNumPages{\pasj}{70}{2}{12}.
\newblock
\begin{APACrefDOI} \doi{10.1093/pasj/psx156} \end{APACrefDOI}
\PrintBackRefs{\CurrentBib}

\bibitem [\protect \citeauthoryear {%
{Kozmanyan}%
, {Bourdin}%
, {Mazzotta}%
, {Rasia}%
\BCBL {}\ \BBA {} {Sereno}%
}{%
{Kozmanyan}%
\ \protect \BOthers {.}}{%
{\protect \APACyear {2019}}%
}]{%
kozmanyan19}
\APACinsertmetastar {%
kozmanyan19}%
\begin{APACrefauthors}%
{Kozmanyan}, A.%
, {Bourdin}, H.%
, {Mazzotta}, P.%
, {Rasia}, E.%
\BCBL {}\ \BBA {} {Sereno}, M.%
\end{APACrefauthors}%
\unskip\
\newblock
\APACrefYearMonthDay{2019}{Jan}{},
\newblock
\unskip
\newblock
\APACjournalVolNumPages{\aap}{621}{}{A34}.
\newblock
\begin{APACrefDOI} \doi{10.1051/0004-6361/201833879} \end{APACrefDOI}
\PrintBackRefs{\CurrentBib}

\bibitem [\protect \citeauthoryear {%
{Markevitch}%
}{%
{Markevitch}%
}{%
{\protect \APACyear {2007}}%
}]{%
markevitch07}
\APACinsertmetastar {%
markevitch07}%
\begin{APACrefauthors}%
{Markevitch}, M.%
\end{APACrefauthors}%
\unskip\
\newblock
\APACrefYearMonthDay{2007}{May}{},
\newblock
\unskip
\newblock
\APACjournalVolNumPages{arXiv e-prints}{}{}{arXiv:0705.3289}.
\PrintBackRefs{\CurrentBib}

\bibitem [\protect \citeauthoryear {%
{Medvedev}%
, {Gilfanov}%
, {Sazonov}%
\BCBL {}\ \BBA {} {Shtykovskiy}%
}{%
{Medvedev}%
\ \protect \BOthers {.}}{%
{\protect \APACyear {2014}}%
}]{%
medvedev14}
\APACinsertmetastar {%
medvedev14}%
\begin{APACrefauthors}%
{Medvedev}, P.%
, {Gilfanov}, M.%
, {Sazonov}, S.%
\BCBL {}\ \BBA {} {Shtykovskiy}, P.%
\end{APACrefauthors}%
\unskip\
\newblock
\APACrefYearMonthDay{2014}{May}{},
\newblock
\unskip
\newblock
\APACjournalVolNumPages{\mnras}{440}{3}{2464-2473}.
\newblock
\begin{APACrefDOI} \doi{10.1093/mnras/stu434} \end{APACrefDOI}
\PrintBackRefs{\CurrentBib}

\bibitem [\protect \citeauthoryear {%
{Nagai}%
\ \BBA {} {Lau}%
}{%
{Nagai}%
\ \BBA {} {Lau}%
}{%
{\protect \APACyear {2011}}%
}]{%
nagai+11}
\APACinsertmetastar {%
nagai+11}%
\begin{APACrefauthors}%
{Nagai}, D.%
\BCBT {}\ \BBA {} {Lau}, E\BPBI T.%
\end{APACrefauthors}%
\unskip\
\newblock
\APACrefYearMonthDay{2011}{{\APACmonth{04}}}{},
\newblock
\unskip
\newblock
\APACjournalVolNumPages{\apjl}{731}{}{L10}.
\newblock
\begin{APACrefDOI} \doi{10.1088/2041-8205/731/1/L10} \end{APACrefDOI}
\PrintBackRefs{\CurrentBib}

\bibitem [\protect \citeauthoryear {%
{Pitrou}%
, {Coc}%
, {Uzan}%
\BCBL {}\ \BBA {} {Vangioni}%
}{%
{Pitrou}%
\ \protect \BOthers {.}}{%
{\protect \APACyear {2018}}%
}]{%
pitrou18}
\APACinsertmetastar {%
pitrou18}%
\begin{APACrefauthors}%
{Pitrou}, C.%
, {Coc}, A.%
, {Uzan}, J\BHBI P.%
\BCBL {}\ \BBA {} {Vangioni}, E.%
\end{APACrefauthors}%
\unskip\
\newblock
\APACrefYearMonthDay{2018}{Sep}{},
\newblock
\unskip
\newblock
\APACjournalVolNumPages{\physrep}{754}{}{1-66}.
\newblock
\begin{APACrefDOI} \doi{10.1016/j.physrep.2018.04.005} \end{APACrefDOI}
\PrintBackRefs{\CurrentBib}

\bibitem [\protect \citeauthoryear {%
{Planck Collaboration}%
\ \protect \BOthers {.}}{%
{Planck Collaboration}%
\ \protect \BOthers {.}}{%
{\protect \APACyear {2014}}%
}]{%
SZcatalog}
\APACinsertmetastar {%
SZcatalog}%
\begin{APACrefauthors}%
{Planck Collaboration}%
, {Ade}, P\BPBI A\BPBI R.%
, {Aghanim}, N.%
\ et al.\end{APACrefauthors}%
\unskip\
\newblock
\APACrefYearMonthDay{2014}{{\APACmonth{11}}}{},
\newblock
\unskip
\newblock
\APACjournalVolNumPages{\aap}{571}{}{A29}.
\newblock
\begin{APACrefDOI} \doi{10.1051/0004-6361/201321523} \end{APACrefDOI}
\PrintBackRefs{\CurrentBib}

\bibitem [\protect \citeauthoryear {%
{Planck Collaboration}%
\ \protect \BOthers {.}}{%
{Planck Collaboration}%
\ \protect \BOthers {.}}{%
{\protect \APACyear {2018}}%
}]{%
planck18-6}
\APACinsertmetastar {%
planck18-6}%
\begin{APACrefauthors}%
{Planck Collaboration}%
, {Aghanim}, N.%
, {Akrami}, Y.%
\ et al.\end{APACrefauthors}%
\unskip\
\newblock
\APACrefYearMonthDay{2018}{Jul}{},
\newblock
\unskip
\newblock
\APACjournalVolNumPages{arXiv e-prints}{}{}{arXiv:1807.06209}.
\PrintBackRefs{\CurrentBib}

\bibitem [\protect \citeauthoryear {%
{Qin}%
\ \BBA {} {Wu}%
}{%
{Qin}%
\ \BBA {} {Wu}%
}{%
{\protect \APACyear {2000}}%
}]{%
qin00}
\APACinsertmetastar {%
qin00}%
\begin{APACrefauthors}%
{Qin}, B.%
\BCBT {}\ \BBA {} {Wu}, X\BHBI P.%
\end{APACrefauthors}%
\unskip\
\newblock
\APACrefYearMonthDay{2000}{Jan}{},
\newblock
\unskip
\newblock
\APACjournalVolNumPages{\apjl}{529}{1}{L1-L4}.
\newblock
\begin{APACrefDOI} \doi{10.1086/312445} \end{APACrefDOI}
\PrintBackRefs{\CurrentBib}

\bibitem [\protect \citeauthoryear {%
{Riess}%
, {Casertano}%
, {Yuan}%
, {Macri}%
\BCBL {}\ \BBA {} {Scolnic}%
}{%
{Riess}%
\ \protect \BOthers {.}}{%
{\protect \APACyear {2019}}%
}]{%
riess19}
\APACinsertmetastar {%
riess19}%
\begin{APACrefauthors}%
{Riess}, A\BPBI G.%
, {Casertano}, S.%
, {Yuan}, W.%
, {Macri}, L\BPBI M.%
\BCBL {}\ \BBA {} {Scolnic}, D.%
\end{APACrefauthors}%
\unskip\
\newblock
\APACrefYearMonthDay{2019}{May}{},
\newblock
\unskip
\newblock
\APACjournalVolNumPages{\apj}{876}{1}{85}.
\newblock
\begin{APACrefDOI} \doi{10.3847/1538-4357/ab1422} \end{APACrefDOI}
\PrintBackRefs{\CurrentBib}

\bibitem [\protect \citeauthoryear {%
{Roncarelli}%
\ \protect \BOthers {.}}{%
{Roncarelli}%
\ \protect \BOthers {.}}{%
{\protect \APACyear {2013}}%
}]{%
roncarelli+13}
\APACinsertmetastar {%
roncarelli+13}%
\begin{APACrefauthors}%
{Roncarelli}, M.%
, {Ettori}, S.%
, {Borgani}, S.%
, {Dolag}, K.%
, {Fabjan}, D.%
\BCBL {}\ \BBA {} {Moscardini}, L.%
\end{APACrefauthors}%
\unskip\
\newblock
\APACrefYearMonthDay{2013}{{\APACmonth{07}}}{},
\newblock
\unskip
\newblock
\APACjournalVolNumPages{\mnras}{432}{}{3030-3046}.
\newblock
\begin{APACrefDOI} \doi{10.1093/mnras/stt654} \end{APACrefDOI}
\PrintBackRefs{\CurrentBib}

\bibitem [\protect \citeauthoryear {%
{Sanders}%
\ \BBA {} {Fabian}%
}{%
{Sanders}%
\ \BBA {} {Fabian}%
}{%
{\protect \APACyear {2013}}%
}]{%
sanders13}
\APACinsertmetastar {%
sanders13}%
\begin{APACrefauthors}%
{Sanders}, J\BPBI S.%
\BCBT {}\ \BBA {} {Fabian}, A\BPBI C.%
\end{APACrefauthors}%
\unskip\
\newblock
\APACrefYearMonthDay{2013}{Mar}{},
\newblock
\unskip
\newblock
\APACjournalVolNumPages{\mnras}{429}{3}{2727-2738}.
\newblock
\begin{APACrefDOI} \doi{10.1093/mnras/sts543} \end{APACrefDOI}
\PrintBackRefs{\CurrentBib}

\bibitem [\protect \citeauthoryear {%
{Silk}%
}{%
{Silk}%
}{%
{\protect \APACyear {1968}}%
}]{%
silk68}
\APACinsertmetastar {%
silk68}%
\begin{APACrefauthors}%
{Silk}, J.%
\end{APACrefauthors}%
\unskip\
\newblock
\APACrefYearMonthDay{1968}{Feb}{},
\newblock
\unskip
\newblock
\APACjournalVolNumPages{\apj}{151}{}{459}.
\newblock
\begin{APACrefDOI} \doi{10.1086/149449} \end{APACrefDOI}
\PrintBackRefs{\CurrentBib}

\bibitem [\protect \citeauthoryear {%
{Silk}%
\ \BBA {} {White}%
}{%
{Silk}%
\ \BBA {} {White}%
}{%
{\protect \APACyear {1978}}%
}]{%
silk78}
\APACinsertmetastar {%
silk78}%
\begin{APACrefauthors}%
{Silk}, J.%
\BCBT {}\ \BBA {} {White}, S\BPBI D\BPBI M.%
\end{APACrefauthors}%
\unskip\
\newblock
\APACrefYearMonthDay{1978}{Dec}{},
\newblock
\unskip
\newblock
\APACjournalVolNumPages{\apjl}{226}{}{L103-L106}.
\newblock
\begin{APACrefDOI} \doi{10.1086/182841} \end{APACrefDOI}
\PrintBackRefs{\CurrentBib}

\bibitem [\protect \citeauthoryear {%
{Spitzer}%
}{%
{Spitzer}%
}{%
{\protect \APACyear {1956}}%
}]{%
spitzer56}
\APACinsertmetastar {%
spitzer56}%
\begin{APACrefauthors}%
{Spitzer}, L.%
\end{APACrefauthors}%
\unskip\
\newblock
\APACrefYear{1956},
\newblock
\APACrefbtitle {{Physics of Fully Ionized Gases}} {{Physics of Fully Ionized
  Gases}}.
\newblock
\APACaddressPublisher{}{New York: Interscience Publishers}.
\PrintBackRefs{\CurrentBib}

\bibitem [\protect \citeauthoryear {%
{Sunyaev}%
\ \BBA {} {Zeldovich}%
}{%
{Sunyaev}%
\ \BBA {} {Zeldovich}%
}{%
{\protect \APACyear {1972}}%
}]{%
SZ}
\APACinsertmetastar {%
SZ}%
\begin{APACrefauthors}%
{Sunyaev}, R\BPBI A.%
\BCBT {}\ \BBA {} {Zeldovich}, Y\BPBI B.%
\end{APACrefauthors}%
\unskip\
\newblock
\APACrefYearMonthDay{1972}{{\APACmonth{11}}}{},
\newblock
\unskip
\newblock
\APACjournalVolNumPages{Comments on Astrophysics and Space Physics}{4}{}{173}.
\PrintBackRefs{\CurrentBib}

\bibitem [\protect \citeauthoryear {%
{Verde}%
, {Treu}%
\BCBL {}\ \BBA {} {Riess}%
}{%
{Verde}%
\ \protect \BOthers {.}}{%
{\protect \APACyear {2019}}%
}]{%
verde19}
\APACinsertmetastar {%
verde19}%
\begin{APACrefauthors}%
{Verde}, L.%
, {Treu}, T.%
\BCBL {}\ \BBA {} {Riess}, A\BPBI G.%
\end{APACrefauthors}%
\unskip\
\newblock
\APACrefYearMonthDay{2019}{Jul}{},
\newblock
\unskip
\newblock
\APACjournalVolNumPages{arXiv e-prints}{}{}{arXiv:1907.10625}.
\PrintBackRefs{\CurrentBib}

\bibitem [\protect \citeauthoryear {%
{Wong}%
\ \protect \BOthers {.}}{%
{Wong}%
\ \protect \BOthers {.}}{%
{\protect \APACyear {2019}}%
}]{%
wong19}
\APACinsertmetastar {%
wong19}%
\begin{APACrefauthors}%
{Wong}, K\BPBI C.%
, {Suyu}, S\BPBI H.%
, {Chen}, G\BPBI C\BPBI F.%
\ et al.\end{APACrefauthors}%
\unskip\
\newblock
\APACrefYearMonthDay{2019}{Jul}{},
\newblock
\unskip
\newblock
\APACjournalVolNumPages{arXiv e-prints}{}{}{arXiv:1907.04869}.
\PrintBackRefs{\CurrentBib}

\bibitem [\protect \citeauthoryear {%
{Zhuravleva}%
\ \protect \BOthers {.}}{%
{Zhuravleva}%
\ \protect \BOthers {.}}{%
{\protect \APACyear {2013}}%
}]{%
zhu13}
\APACinsertmetastar {%
zhu13}%
\begin{APACrefauthors}%
{Zhuravleva}, I.%
, {Churazov}, E.%
, {Kravtsov}, A.%
, {Lau}, E\BPBI T.%
, {Nagai}, D.%
\BCBL {}\ \BBA {} {Sunyaev}, R.%
\end{APACrefauthors}%
\unskip\
\newblock
\APACrefYearMonthDay{2013}{{\APACmonth{02}}}{},
\newblock
\unskip
\newblock
\APACjournalVolNumPages{\mnras}{428}{}{3274-3287}.
\newblock
\begin{APACrefDOI} \doi{10.1093/mnras/sts275} \end{APACrefDOI}
\PrintBackRefs{\CurrentBib}

\end{thebibliography}

\end{document}